%% file: template.tex
\documentclass[a4paper]{article}

\usepackage{INTERSPEECH2020}
\usepackage{multirow}
\usepackage{caption}
\usepackage{subcaption}
\usepackage[table,xcdraw]{xcolor}
\usepackage[normalem]{ulem}
\useunder{\uline}{\ul}{}
\usepackage{threeparttable}
\usepackage[export]{adjustbox}

\title{SADDEL: Joint Speech Separation and Denoising Model \\ based on Multitask Learning}
\name{Yuan-Kuei Wu$^1$$^\star$ \qquad Chao-I Tuan$^2$$^\star$ \qquad Hung-Yi Lee$^1$ \qquad Yu Tsao$^3$
\thanks{$^\star$The two first authors made equal contributions} }
\address{
  $^1$Graduate Institute of Communication Engineering, National Taiwan University\\
  $^2$Data Science Degree Program, National Taiwan University\\
  $^3$Research Center for Information Technology Innovation, Academia Sinica}
\email{\{ywk991112, chaoi111.t, tlkagkb93901106\}@gmail.com, yu.tsao@citi.sinica.edu.tw}

\begin{document}

\maketitle
\begin{abstract}
Speech data collected in real-world scenarios often encounters two issues. First, multiple sources may exist simultaneously, and the number of sources may vary with time. Second, the existence of background noise in recording is inevitable. To handle the first issue, we refer to speech separation approaches, that separate speech from an unknown number of speakers. To address the second issue, we refer to speech denoising approaches, which remove noise components and retrieve pure speech signals. Numerous deep learning based methods for speech separation and denoising have been proposed that show promising results. However, few works attempt to address the issues simultaneously, despite speech separation and denoising tasks having similar nature. In this study, we propose a joint speech separation and denoising framework based on the multitask learning criterion to tackle the two issues simultaneously. The experimental results show that the proposed framework not only performs well on both speech separation and denoising tasks, but also outperforms related methods in most conditions.

\end{abstract}

\section{Introduction}


In recent years, deep learning has made substantial progress in speech separation (SS) \cite{YLiu2019CASA,YLuo2019ConvTasnet, wang2014training} and speech denoising (SD) tasks \cite{lu2013speech, liu2014experiments, xu2015regression}. 
By using high-performance SS and SD approaches as front-end units, the performances of automatic speech recognition (ASR) \cite{wang2016joint, li2015robust}, speaker recognition \cite{michelsanti2017conditional,shon2019voiceid} and emotion recognition \cite{emotion1, emotion2} have been improved considerably. However, in real-world scenarios, multiple sources and background noise often occur at the same time. Thus, deriving a unified SS and SD framework that can address the issues simultaneously is an important task and with practical applications. 

A major challenge in combining SS and SD tasks into a unified framework is that noise signals have diverse and unpredictable patterns. For example, unwanted environmental sounds, machine sounds, traffic sounds, etc. have rather different patterns, but they are all considered noise when the target is human speech.

One approach is to cascade SD and SS to first remove noise components, and then perform speech separation \cite{9054572,du2016ustc,wang2020study,ma2020two}. Such cascade approaches have proven effective and exhibit notable improvements in ASR performance under challenging conditions \cite{du2016ustc,wang2020study}.      


In contrast to the cascade approach, we propose a joint SS and SD framework based on the multitask learning criterion named separation and denoising model (SADDEL). The main concept of multitask learning is to simultaneously process two tasks that have shared properties using a unified model \cite{caruana1997multitask}. We argue that although the SS and SD tasks aim to tackle different issues, they share the same intrinsic concept of extracting specific signals from a mixed one. Therefore, the multitask learning criterion is specially suitable.  
Moreover, SADDEL adopts recursive separation, and thus, is able to carry out SD (separating speech and noise) and SS (separating distinct sources from a mixture) without knowing the exact number of source. The experimental results show that SADDEL achieves better separation and denoising as compared to individual SS and SD approaches and presents high robustness among different datasets. 
To the best of our knowledge, this is the first method that combines the SS and SD tasks based on the multitask learning criterion.
As compared to previous methods that may have imbalanced performance (poor separation or denoising results \cite{luo2020separating} \cite{kavalerov2019universal}), SADDEL can perform well without performance sacrifice in individual tasks. 





\section{Related Work}
In this section, we discuss related works that unify the SS and SD tasks into a single framework. A popular approach is to directly cascade the SD and SS models \cite{9054572,du2016ustc,wang2020study,ma2020two}, where the input speech is first processed by an SD model to remove noise components. The denoised signals are considered as having multiple sources, and subsequently processed by an SS model to separate the individual sources. Another approach is building a unified model to perform both tasks using one model  \cite{gao2017unified}. However, the models are usually speaker-dependent, meaning that a target speaker must be specified beforehand, and speech from other speakers are considered as noise. SADDEL is based on a similar concept of the unified model approach, but it does not require a target speaker or knowledge of the number of sources.

Multi-scenario training was first introduced in \cite{zegers2018multi}, where it was proven to be beneficial in training scenarios with multiple sources.  Subsequent studies \cite{luo2020separating} have adopted the framework in \cite{zegers2018multi} and achieved improved SS and SD performance. Kavale rov \cite{kavalerov2019universal} proposed universal SS, which incorporates different window sizes and input feature designs into a separation module for speech and arbitrary sound separation. The proposed framework differs from previous methods in the following aspects. First, environmental noise is introduced in our data 
to simulate real world situations. Second, \cite{kavalerov2019universal} treats SD as an SS task, and considers noise as speech sound in order to unify two tasks. In our work, we assume the noise to be unpredictable and without any learnable pattern. This may be harmful to the separator module and lead to confusing results. Therefore, we introduced keep the SD task in the training scheme to tackle the universal sound separation problem. In other words, only the speech signal is separated in the SD task. 

We adopt the recursive separation model from \cite{takahashi2019recursive} as the main separator module. Based on the recent success of the recursive model in separating varying numbers of speech sources \cite{kavalerov2019universal,takahashi2019recursive,9053921}, we extend this idea to tackle the universal separation problem. The main difference between our work and \cite{takahashi2019recursive} is that we combine speech denoising task into the speech separation task in the training stage.
In addition to the recursive separation model, another stream of approaches to tackle various number of sources is assuming a maximum number. A fault tolerance mechanism is proposed to alleviate misleading training by wrong estimations of actual speaker number in \cite{luo2020separating}. Our preliminary experiments show a notable limitation of this stream of approaches, i.e., the performance in single speaker denoising task is significantly lower than the performance of existing denoising methods. Moreover, another drawback of these approaches is the requirement of setting a maximum speaker number of speakers. The classifying performance may be dynamically affected by such an upper bound.  



\begin{figure}[t]
    \flushleft
    \includegraphics[width=0.9\linewidth]{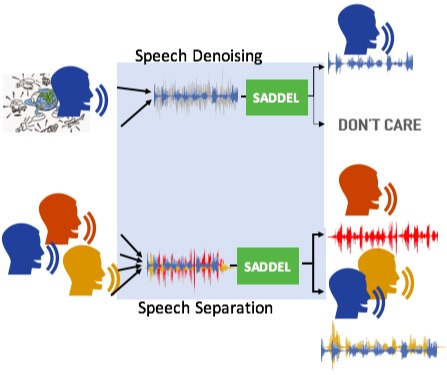}
    \caption{Our paradigm for universal separation via Multitask Learning method}
    \label{fig:model}
\end{figure}

\section{Proposed SADDEL Approach}

The main objective of the multitask methodology \cite{caruana1997multitask} is to train a single network to solve a one task in parallel with other auxiliary tasks. Notably, sharing information among several tasks leads to better performance compared to processing each task independently. Multitask training has been long explored in ASR, speaker adaptation, etc. Nevertheless, its applicability has the following constraint: joint training tasks share at least one degree of similarity. 
We infer that SS and SD are similar in nature and perfectly suitable to be used as paired tasks for multitask learning.

The proposed joint SS and SD framework based on multitasking training to tackle the universal separation problem is Figure \ref{fig:model}. To the best of our knowledge, this is the first work that combines these two tasks via multitask learning. The overall architecture of SADDEL consists of two parts: The first part is the basis transform module that encodes and decodes the sound signals, and the second part is a separation module that is formed by stacked blocks of dilated convolutions similar to the one in \cite{YLuo2019ConvTasnet}. All networks are trained using the negative scale-invariant signal-to-noise ratio (SI-SNR) \cite{le2019sdr} as the loss function between the outputs and targets. 
We consider the SS and SD scenarios to achieve universal source separation. For the SS scenario, SADDEL aims to extract pure speech sources from a 2-speaker mixture and 3-speaker mixture, with and without noise components. For the SD scenario, SADDEL aims to extract pure speech from a 1-speaker source with noise components. The data for the SS and SD scenarios are combined to form the training set. 
During training, speech utterances from the training set are randomly sampled to form a batch. For each step, the model run over separation and denoising tasks (e.g., separation on a 2-speaker mixture and 3-speaker mixture with or without noise, and denoising on 1-speaker source with noise components), and the objective loss is average among all tasks and then update the model once. 

We follow the recursive separation process \cite{takahashi2019recursive} to apply our framework for separating varying number of speakers. In single channel SS, we aim to separate N speaker sources $s_1(t), s_2(t), \cdots, s_N(t)$ from the mixture signal $m(t)$, where $m(t)$ can be denoted as $m(t) = \sum_{i=1}^{N}{s_i (t)}$. Iterative separation separates a single desired source at a time, and the remaining mixture becomes the input of the next separation process. At the $j$-th recursive step, our model generates two separated outputs $\hat{s}^j(t)$ and $\hat{r}^j(t)$. The first output, $\hat{s}^j(t)$, should be close to one speaker source in $\hat{r}^{j-1}(t)$, and the second output, $\hat{r}^j(t)$, should be the remainder of $\hat{r}^{j-1}(t)$. This is formulated as:
\begin{equation}
    \hat{s}^j(t), \hat{r}^{j}(t) =  F(\hat{r}^{j-1}(t)),
\end{equation}
where $F$ denotes the separation model. Permutation invariant training (PIT) \cite{pit_multitalker,pit} is generally used to solve the label permutation problem in speaker-independent SS. Instead of generating masks for specific speakers, PIT calculates the loses for all ($N!$) possible permutations of the speakers in the outputs, and selects the order that generates the minimum loss as the label assignment. However, in our problem setting, only one speaker source is separated from the remainder of the mixture. Consequently, there are only $N$ possible permutations rather than $N!$ (i.e., only $N$ sources can be separated at a time).

We adopted the one-and-rest PIT (OR-PIT) \cite{takahashi2019recursive} to train the separation module in SADDEL. 
OR-PIT was proposed to solve a multiple source problem by reforming the conventional PIT calculation into one and the rest assignment. We argue that the SI-SNR loss won't be scaled by the number of speakers in the source due to the scale invariant characteristic. Therefore, we modify the equation and our objective function is formulated as,
\begin{equation} \label{eq:obj1}
    L = \min_i l(\hat{s}(t), s_i(t)) + l(\hat{r}(t), \sum_{n \neq i}(s_n(t))
\end{equation}
where $l(\cdot )$ denotes SADDEL.



\section{Experiments}
In this section, we first introduce the datasets for evaluation, and then present the experimental setup and finally the results. 
\subsection{Experimental Setup}
\subsubsection{Dataset}
Four datasets were used to evaluate the proposed SADDEL approach: \textbf{WSJ0-mix}, \textbf{MUSAN}, \textbf{WHAM!}, \textbf{100 Nonspeech}, and \textbf{Librispeech-mix}. In the following, we will discuss the details of these datasets.\\
\textbf{WSJ0-mix} We train and evaluate our proposed method on the publicly available datasets WSJ0-2mix and WSJ0-3mix \cite{deep_cluster}, which were derived from the WSJ0 corpus. From the si\_tr\_s dataset, 30 hours of training data and 10 hours of validation data were generated. Following the approach in \cite{deep_cluster,takahashi2019recursive}, the speech utterances were randomly sampled and mixed up between SNR from -2.5 dB to 2.5dB. All waveforms were downsampled to 8KHz in the pre-processing step.\\ 
\textbf{MUSAN} The noisy data used in for training and testing comes from the MUSAN dataset \cite{Musna2015}, which includes various technical and non-technical noises profiles such as DTMF tones, dial tones, and fax machine noises, are included. Additionally, ambient sounds such as idling cars, thunder, and animal noises, are included. Although previous works \cite{kavalerov2019universal,9053921} exclude ambience and environmental sounds when considering the universal separation problem, we believe they are present in most real-world conditions. The noise was sampled between SNR -5 dB and 20 dB, and mixed for training.\\
\textbf{WHAM!} We perform testing experiments on the noisy data recorded in public areas by \cite{wichern2019wham}, called WHAM!.\\
\textbf{100 Nonspeech} This is a collection of 100 sound tracks of 20 different types of non-speech sounds \cite{100nonspeech2010}.
, which is used in our testing.\\  
\textbf{Librispeech-mix} We simulate another single-channel SS dataset for testing with the Librispeech dataset \cite{panayotov2015librispeech}, where test data is generated from a 100-hour test set. All utterances are 6 seconds long with a sample rate of 8 KHz.

\subsubsection{Implementation Details}
Our methods can be applied to any type of separation model, so the main focus of the present study is to verify the effectiveness of the multitask learning criteria, rather than comparing different model types. The ConvTasnet \cite{YLuo2019ConvTasnet} model has shown to provide promising results for both SS and SD tasks, and thus, it was selected to build the separation module in SADDEL.   
%

All of the models presented in this work were implemented in Pytorch. Each model was trained for 100 epochs on 4 parallel NVIDIA Tesla v100 GPUs, using the Adam optimizer. We followed the best configurations for ConvTasnet \cite{YLuo2019ConvTasnet} to build the separator module. Our initial learning rate was set to $1e-3$, and the weight decay was applied thereafter. The performance was measured in SI-SNRi \cite{measurement2006}. 

\subsubsection{Comparative Model I: Cascade Learning}

In this study, we implemented a cascade learning model as a comparative approach. To train this cascade model, we used the noisy 2-speaker and noisy 3-speaker as the training dataset. In the first stage, SD is applied to the noisy mixture, which is then used as the input for the SS step in the second stage.
The SD and SS modules are simultaneously trained to minimize the overall loss.

\subsubsection{Comparative Model II: Auxiliary Autoencoding PIT}
In addition to the cascade learning approach, we prepared another comparative model with the setting proposed in \cite{luo2020separating}.
This method separates varying numbers of speakers in a mixture with "fault tolerance" ability. Aside from the recursive methods, they follow the "fix-output-number" method proposed in \cite{pit_multitalker, DA2018}, where a maximum number of speakers is required in advance. The main objective is to add autoencoding in the PIT. Since the number of the speaker sources in the mixture is possibly smaller than the maximum number setting, the input mixture speech is set to be the output target for the redundant output channel, which performs a null separation process. 
To achieve a reasonable comparison, we implement the autoencoding setting in our experiment under the same separation module and other environmental condition. 


\begin{table}[t]
\centering
\caption{Average SI-SNRi scores on SD, SS without noise (CSS) and SS under noise (NSS) between SADDEL and the comparative models. We colored the results in gray to denote which specific configuration was used in training phase.}
\label{table:result}
\input{result_table}
\end{table}

\subsection{Experimental Results}
 There are five training and testing set configurations in our experiments: one speaker source with real-world noise (SD scenario, denoted as 1sp+n), multiple speakers without noise (SS in clean scenario, denoted as 2sp and 3sp for 2-speaker and 3-speaker mixtures, respectively), and multiple speakers with noise (SS in noise scenario, denoted as 2sp+n and 3sp+n for 2-speaker and 3-speaker mixtures in noise, respectively). The three comparative models are the cascade model (termed cascade), conventional ConvTasnet models(used to build baseline\_SS and baseline\_SD), and the auxiliary autoencoding PIT ConvTasnet model (termed A2PIT). SADDEL used the combined \{1sp+n, 2sp, 3sp, 2sp+n, and 3sp+n\} training set, and the baseline\_SD model (denoted as bl\_SD) was trained with the 1sp+n set. Moreover, the baseline\_SS model (denoted as bl\_SS) was trained with the combined \{2sp+n and 3sp+n\} set. For A2PIT, we used the combined \{1sp+n, 2sp+n and 3sp+n\} set, and for cascade, we used the combined \{2sp+n and 3sp+n\} set. 

Table \ref{table:result} compares our multitask learning model with other comparative models on the WSJ0-mix dataset. We note that SADDEL outperforms comparative systems across all testing sets, indicating that training with data in all tasks improves generalization by sharing useful information for the SD and SS tasks. 
Moreover, we note that the number of parameters of SADDEL is only half to the cascade model while still achieves better performance. This is advantageous in integrating n additional advantage that the proposed model SADDEL can be more suitably installed with in edge devices.
It has been noted in \cite{luo2020separating} that A2PIT performs poorly in the single speaker denoising task. In our opinion, selecting invalid estimated sources due to a wrongly predicted speaker count causes performance degradation in the single speaker denoising task. SADDEL has better performance in one speaker denoising tasks, and separation tasks with multiple speakers. 

\begin{table}[t]
\centering
\caption{Average SI-SNRi scores of SADDEL and comparative approahces on Librispeech-mix (denoted as L) and WHAM! (denoted as W).}
\label{table:dif_dataset}
\begin{threeparttable}
\input{dataset_table}

\end{threeparttable}
\end{table}

\begin{table}[t]
\centering
\caption{Average SI-SNRi scores of SADDEL and comparative approaches on Librispeech-mix (denoted as L) and wsj0 with different SNR levels, where \text{M\_C} denotes Musan dataset at middle SNR level (SNR= -5 to - 25 dB), \text{M\_L} denotes Musan dataset at low SNR level (SNR= -5 to - 5 dB). and \text{N\_H} denotes  100Nonspeech dataset with high SNR level (SNR= 10 to 20 dB)}
\label{table:vary_noise}
\begin{threeparttable}
\input{ndataset_table}
\footnotesize

\end{threeparttable}
\end{table}

\subsection{Robustness}

To further examine the performance robustness, we tested SADDEL in different scenarios. In Table \ref{table:dif_dataset}, we present the results of SADDEL and the comparative models on two testing datasets (formed by Librispeech-mix and WHAM!). The test set in Librispeech-mix is mixed with random sampled noise from the MUSAN dataset at the SNR level for training. Additionally, WHAM! is already a noisy SS task and requires no such additional mixing. As shown in Table \ref{table:dif_dataset}, SADDEL outperforms almost all the comparative approaches, but it slightly underperforms the baseline\_SD model in the SD task. Moreover for the WHAM! test set, SADDEL surpasses all the comparative models by nearly one SI-SNR improvement.

In Table \ref{table:vary_noise}, we present the results of SADDEL and comparative models with noise types proposed in \cite{luo2020separating} at three SNR levels. 
From the results in Table \ref{table:vary_noise}, we note that when switching to higher/lower SNR levels and different noise types, SADDEL retains better performance than comparative models, where the SS performance notably dropped when unseen noises are involved. Furthermore, We observed a severe performance drop for A2PIT. A potential explanation is that A2PIT needs to define a threshold to identify a valid speaker count, and an optimal threshold may not be easily defined when unseen noises are involved. 
The comparable performance of the A2PIT model on the WHAM! dataset may be due to the high similarity between the noise data collected from the WHAM! and MUSAN dataset. When testing on unseen test data and noise, SADDEL outperforms all the comparative models in all test conditions. 



\begin{figure}
\flushleft
\caption{Performance degradation between SS without noise and SS with noise on the WSJ0-mix dataset.}
\includegraphics[scale=0.4]{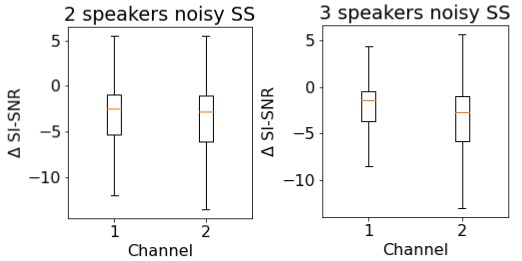}
\label{fig:degradation}
\end{figure}

\subsection{Noise Impact on Channel}
Since the output from the first channel in our model is considered as a pure source, we assume that noise components have been entirely removed. The output of the second channel can have a higher noise tolerance level due to the reduction of noise is able to apply in the following recursive step. To analyze this, we first sent two mixtures (a clean mixture and noisy mixture) into the SADDEL model, and measured the SI-SNR scores of the outputs. We subtract the SI-SNR results channel-wise, and a larger value indicates that the noise causes less impact. Figure 2 shows that the performance degradation in the first channel is less than that in the second channel. Furthermore, the means of SI-SNR degradation are -3.4822 and -3.9743 for the first and second channels, respectively, in the 2-speaker noisy SS, and -2.3738 and -3.8146 for the first and second channels, respectively, in 3-speaker noisy SS. Our recursive SS approach should be benefit from this phenomenon.


\section{Conclusion}
In this study, we propose a novel SADDEL approach to perform noisy SS with an unknown number of speakers. In contrast to existing methods, SADDEL is derived based on the multitask learning criterion, and thus, a unified model is used to carry out SD and SS simultaneously. This design notably reduces the hardware cost as compared to conventional cascade methods. The experimental results demonstrate that SADDEL outperforms comparative SD and SS models, and exhibits promising results on various noisy SS tasks. Moreover, SADDEL can provide high performance robustness across different datasets, noise types, and SNR levels.

\section{Acknowledgements}
We are grateful to the National Center for High-performance Computing for computer time and facilities.

\bibliographystyle{IEEEtran}

\bibliography{mybib}


\end{document}

%% file: result_table.tex
\begin{tabular}{c|c|cc|cc}
                            & \multicolumn{5}{c}{\textbf{SI-SNRi}}                                                                                                                                                                     \\ \cline{2-6} 
                            & \textbf{SD}                            & \multicolumn{2}{c|}{\textbf{CSS}}                                               & \multicolumn{2}{c}{\textbf{NSS}}                      \\ \cline{2-6} 
\multirow{-3}{*}{\textbf{}} & \textbf{1sp+n}                         & \textbf{2sp}                           & \textbf{3sp}                           & \textbf{2sp+n}                         & \textbf{3sp+n}                        \\ \hline
\textbf{SADDEL}              & \cellcolor[HTML]{EFEFEF}\textbf{10.74} & \cellcolor[HTML]{EFEFEF}\textbf{15.22} & \cellcolor[HTML]{EFEFEF}\textbf{11.95} & \cellcolor[HTML]{EFEFEF}\textbf{14.31} & \cellcolor[HTML]{EFEFEF}\textbf{11.6} \\ \hline
\textbf{bl\_SS}             & \cellcolor[HTML]{FFFFFF}0.15           & \cellcolor[HTML]{FFFFFF}12.99          & \cellcolor[HTML]{FFFFFF}10.82          & \cellcolor[HTML]{EFEFEF}12.83          & \cellcolor[HTML]{EFEFEF}11.14         \\ \hline
\textbf{bl\_SD}             & \cellcolor[HTML]{EFEFEF}9.9            & \cellcolor[HTML]{FFFFFF}-1.48          & \cellcolor[HTML]{FFFFFF}-1.97          & \cellcolor[HTML]{FFFFFF}0.38           & \cellcolor[HTML]{FFFFFF}-0.3          \\ \hline
\textbf{A2PIT}              & \cellcolor[HTML]{EFEFEF}8.35           & \cellcolor[HTML]{FFFFFF}13.04          & \cellcolor[HTML]{FFFFFF}10.23          & \cellcolor[HTML]{EFEFEF}12.79          & \cellcolor[HTML]{EFEFEF}10.5          \\ \hline
\textbf{cascade}            & \cellcolor[HTML]{FFFFFF}7.99           & \cellcolor[HTML]{FFFFFF}13.1           & \cellcolor[HTML]{FFFFFF}10.13          & \cellcolor[HTML]{EFEFEF}13.44          & \cellcolor[HTML]{EFEFEF}10.67        
\end{tabular}

%% file: dataset_table.tex
\begin{tabular}{cl|c|cc|cc}
                                   &    & \multicolumn{5}{c}{\textbf{SI-SNRi}}                                                                                                                                                                                                                                                                    \\ \cline{3-7} 
                                   &    & \textbf{SD}                                                  & \multicolumn{2}{c|}{\textbf{CSS}}                                                                                   & \multicolumn{2}{c}{\textbf{NSS}}                                                                                   \\ \cline{3-7} 
\multirow{-3}{*}{\textbf{}}        &    & \textbf{1sp+n}                                               & \textbf{2sp}                                                 & \textbf{3sp}                                         & \textbf{2sp+n}                                               & \textbf{3sp+n}                                       \\ \hline
                                   & L\tnote{*} & \cellcolor[HTML]{EFEFEF}{\color[HTML]{000000} 4.51}          & \cellcolor[HTML]{EFEFEF}{\color[HTML]{000000} \textbf{9.33}} & \cellcolor[HTML]{EFEFEF}{\color[HTML]{000000} \textbf{7.18}}  & \cellcolor[HTML]{EFEFEF}{\color[HTML]{000000} \textbf{9.69}} & \cellcolor[HTML]{EFEFEF}{\color[HTML]{000000} \textbf{7.99}}  \\
\multirow{-2}{*}{\textbf{SADDEL}}   & W\tnote{*} & -                                                            & -                                                            & \textbf{-}                                           & 11.97                                                        & -                                                    \\ \hline
                                   & L  & \cellcolor[HTML]{EFEFEF}{\color[HTML]{000000} -1.90}         & \cellcolor[HTML]{EFEFEF}8.28                                 & \cellcolor[HTML]{EFEFEF}6.18                         & \cellcolor[HTML]{EFEFEF}8.91                                 & \cellcolor[HTML]{EFEFEF}7.23                         \\
\multirow{-2}{*}{\textbf{bl\_SS}}  & W  & -                                                            & -                                                            & -                                                    & 6.4                                                          & -                                                    \\
                                   & L  & \cellcolor[HTML]{EFEFEF}{\color[HTML]{000000} \textbf{5.25}} & \cellcolor[HTML]{EFEFEF}{\color[HTML]{000000} -1.24}         & \cellcolor[HTML]{EFEFEF}{\color[HTML]{000000} -1.59} & \cellcolor[HTML]{EFEFEF}{\color[HTML]{000000} 0.32}          & \cellcolor[HTML]{EFEFEF}{\color[HTML]{000000} -0.07} \\
\multirow{-2}{*}{\textbf{bl\_SD}}  & W  & -                                                            & -                                                            & {\color[HTML]{000000} -}                             & -2.96                                                        & -                                                    \\ \hline
                                   & L  & \cellcolor[HTML]{EFEFEF}-1.27                                & \cellcolor[HTML]{EFEFEF}7.79                                 & \cellcolor[HTML]{EFEFEF}6.27                         & \cellcolor[HTML]{EFEFEF}8.94                                 & \cellcolor[HTML]{EFEFEF}7.65                         \\
\multirow{-2}{*}{\textbf{cascade}} & W  & -                                                            & -                                                            & -                                                    & 11.12                                                        & -                                                    \\ \hline
                                   & L  & \cellcolor[HTML]{EFEFEF}3.51                                 & \cellcolor[HTML]{EFEFEF}7.49                                 & \cellcolor[HTML]{EFEFEF}6.53                         & \cellcolor[HTML]{EFEFEF}8.36                                 & \cellcolor[HTML]{EFEFEF}7.34                         \\
\multirow{-2}{*}{\textbf{A2PIT}}   & W  & -                                                            & -                                                            & -                                                    & 10.9                                                         & -                                                   
\end{tabular}

%% file: ndataset_table.tex
\begin{tabular}{lll|l|ll}
                            & \multicolumn{1}{c}{}                                   & \multicolumn{1}{c|}{}                                 & \multicolumn{3}{c}{\textbf{SI-SNRi}}                               \\ \cline{4-6} 
                            & \multicolumn{1}{c}{}                                   & \multicolumn{1}{c|}{}                                 & \multicolumn{1}{c|}{\textbf{SD}} & \multicolumn{2}{c}{\textbf{NSS}} \\ \cline{4-6} 
\multirow{-3}{*}{\textbf{}} & \multicolumn{1}{c}{\multirow{-3}{*}{\textbf{Dataset}}} & \multicolumn{1}{c|}{\multirow{-3}{*}{\textbf{Noise}}} & \textbf{1sp+n}                   & \textbf{2sp+n}  & \textbf{3sp+n} \\ \hline
\rowcolor[HTML]{EFEFEF} 
\textbf{SADDEL}              & wsj0                                                   & M\_C\tnote{*}                                                 & \textbf{10.74}                   & \textbf{14.31}  & \textbf{11.6}  \\ \hline
\textbf{bl\_SS}             & wsj0                                                   & M\_C                                                  & 0.15                             & 12.83           & 11.14          \\
\textbf{bl\_SD}             & wsj0                                                   & M\_C                                                  & 9.9                              & 0.38            & -0.3           \\
\textbf{A2PIT}              & wsj0                                                   & M\_C                                                  & 8.35                             & 12.79           & 10.5           \\
\textbf{cascade}            & wsj0                                                   & M\_C                                                  & 7.99                             & 13.4            & 11.6           \\ \hline
\rowcolor[HTML]{EFEFEF} 
\textbf{SADDEL}              & wsj0                                                   & M\_L\tnote{*}                                                  & \textbf{15}                   & \textbf{14.57}           & \textbf{12.08} \\ \hline
\textbf{A2PIT}              & wsj0                                                   & M\_L                                                  & 13.06                            & 13.58            & 11.3             \\
\textbf{cascade}            & wsj0                                                   & M\_L                                                  & 13.52                             & 14.16   & 11.97  \\ \hline
\rowcolor[HTML]{EFEFEF} 
\textbf{SADDEL}              & wsj0                                                   & N\_H\tnote{*}                                                   & \textbf{8.03}                    & \textbf{13.37}  & \textbf{10.53} \\ \hline
\textbf{bl\_SS}             & wsj0                                                   & N\_H                                                  & -6.19                            & 11.76           & 10             \\
\textbf{bl\_SD}             & wsj0                                                   & N\_H                                                  & 7.5                              & -1.01           & -1.55          \\
\textbf{A2PIT}              & wsj0                                                   & N\_H                                                  & 2.77                             & 11.92           & 9.74           \\
\textbf{cascade}            & wsj0                                                   & N\_H                                                  & 6.37                             & 12.34           & 9.57           \\ \hline
\rowcolor[HTML]{EFEFEF} 
\textbf{SADDEL}              & Libri                                                  & N\_H                                                  & 1.91                    & \textbf{8.58}   & \textbf{6.58}  \\ \hline
\textbf{bl\_SS}             & Libri                                                  & N\_H                                                  & -5.72                            & 7.42            & 5.88           \\
\textbf{bl\_SD}             & Libri                                                  & N\_H                                                  & \textbf{3.56}                    & -0.8            & -1.2           \\
\textbf{A2PIT}              & Libri                                                  & N\_H                                                  & -0.18                            & 7.04            & 6.27           \\
\textbf{cascade}            & Libri                                                  & N\_H                                                  & \cellcolor[HTML]{FFFFFF}-3.11    & 7.87            & 6.42          
\end{tabular}